# Alloy inhomogeneity and carrier localization in AlGaN sections and AlGaN/AlN nanodisks in nanowires with 240-350 nm emission

C. Himwas[1,2], M. den Hertog[1,2], Le Si Dang[1,2], E. Monroy[1,3], and R. Songmuang[1,2]
[1] Université Grenoble Alpes, 38000 Grenoble, France
[2] Institut Néel-CNRS, 25 rue des Martyrs, 38042 Grenoble Cedex 9, France
[3] CEA-Grenoble, INAC/SP2M/NPSC, 17 rue des Martyrs, 38054 Grenoble Cedex 9, France

The Al-Ga intermixing at Al(Ga)N/GaN interfaces in nanowires and the chemical inhomogeneity in $Al_xGa_{1-x}N$/AlN nanodisks (NDs) are attributed to the strain relaxation process. This interpretation is supported by the three-dimensional strain distribution calculated by minimizing the elastic energy in the structure. The alloy inhomogeneity increases with Al content, leading to enhanced carrier localization signatures in their optical characteristics i.e. red shift of the emission, *s*-shaped temperature dependence and linewidth broadening. Despite these alloy fluctuations, the emission energy of AlGaN/AlN NDs can be tuned in the 240-350 nm range with internal quantum efficiencies around 30%.

AlGaN/AlN quantum wells (QWs)[1] and quantum dots (QDs)[2,3,4] have become promising active media for solid-state ultraviolet lighting. Nanodisks (NDs) in nanowires (NWs) have also emerged as alternative nanostructures which offer not only strong carrier localization[5,6] but also a possibility to improve photon extraction efficiency[7]. However, synthesis of ternary alloys in NWs generally suffers from strong chemical inhomogeneities[8], which modify their electronic properties with respect to random alloys.

Here, we present structural and optical studies of AlGaN sections and AlGaN/AlN NDs in NWs. In the case of AlGaN sections, alloy gradient and spontaneous chemical ordering were found along the wire axis, in addition to the formation of an Al-rich shell. We found that alloy fluctuations increase in AlGaN sections with higher Al content, as evidenced by stronger carrier localization in their optical characteristics. At GaN/Al(Ga)N interfaces, we identify a *transition zone* with strong Al-Ga intermixing whose formation is attributed to the strain relaxation process. This interpretation is supported by a three-dimensional (3D) strain simulation. Despite the radial and axial alloy inhomogeneity, the photon emission energy of AlGaN/AlN NDs can be tuned by adjusting the nominal thickness and Al content.

AlGaN sections and stacks of AlGaN/AlN NDs on GaN NWs were grown on Si (111) by plasma-assisted molecular beam epitaxy (PAMBE). Initially, a thin (~2 nm) AlN buffer layer[9] was grown at the substrate temperature $T_s = 840°C$. Then, $T_s$ was decreased to 795 °C for the growth of GaN NWs using a Ga flux $\Phi_{Ga} \sim 0.17$ ML/s while the flux of active nitrogen was set at $\Phi_N = 0.34$ ML/s to maintain N-rich conditions. The deposition time of GaN NWs was $t = 2.5$ hours. Then, the $Al_xGa_{1-x}N$ sections or the $Al_xGa_{1-x}N$/AlN NDs were synthesized on the top of these GaN NWs. The deposition time of AlGaN sections was 40 min, while that of AlGaN NDs was between 10 and 40 s with ~1 min AlN barriers. To adjust the nominal Al content ($x = \Phi_{Al}/\Phi_N$) in the $Al_xGa_{1-x}N$, the Al flux was varied in the $\Phi_{Al} = 0.07$-$0.17$ ML/s range, while keeping $\Phi_{Ga} = 0.09$ ML/s and $\Phi_N = 0.34$ ML/s for the entire growth process. The Al, Ga and N deposition rates were calibrated by reflection high-energy electron diffraction intensity oscillations during the growth of two-dimensional layers of GaN or AlN on GaN/sapphire substrates at 720°C. The substrate temperature for the calibration is lower than the growth temperature of the NWs. Thus, the real deposition rate should be lower than the calibration due to the thermal activation of Ga desorption and GaN decomposition.

The NW morphology was characterized by field emission scanning electron microscopy using a Zeiss Ultra55. The typical NW diameter is around 60 nm, with a length of 600 nm. The polarity of GaN NWs is found to be N-polar[10]. Detailed structural properties were measured by high angle annular dark field scanning transmission electron microscopy (HAADF-STEM) using a probe-corrected FEI Titan operated at 300 kV. Photoluminescence (PL) was performed by exciting the samples with a continuous-wave frequency-doubled $Ar^+$ laser ($\lambda$=244 nm). The excitation power was around 100 µW with a spot diameter of 100 µm. The emission from the sample was collected by a Jobin Yvon HR460 monochromator equipped with a UV-enhanced charge-coupled device (CCD) camera. Cathodoluminescence (CL) measurements were performed in a FEI quanta 200 CL system, using an acceleration voltage of 10 kV and a current of ~100 pA. The sample emission was collected by a parabolic mirror, which focused it onto a CCD camera.

Deduced from HAADF-STEM images of several wires, the average length of the $Al_xGa_{1-x}N$ sections on the GaN NWs was ~180 nm. This value corresponds to an AlGaN vertical growth rate $R_{AlGaN} \sim 0.30$ ML/s[11], independent of $\Phi_{Al}$ and nearly equal to $\Phi_N$, indicating that the AlGaN growth is limited by the N- flux.[12] In addition, $R_{AlGaN}$ is larger than the summation of the Al and Ga fluxes ($\Phi_{Ga}+\Phi_{Al} \leq 0.26$ ML/s). This fact confirms that the adatom diffusion at the NW sidewall contributes to the NW growth, in addition to the direct material deposition on the NW top. The majority of the diffusing adatoms are supposed to be Ga since their diffusion length is longer than that of Al and N adatoms.[12,13]

Figure 1(a) presents an HAADF-STEM image of an $Al_{0.5}Ga_{0.5}N$ section on a GaN NW. The bright contrast is GaN whereas the darker contrast implies higher Al content. Along the radius of the NWs, we observe an Al-rich shell encapsulating both the AlGaN section and GaN NW base, which spontaneously forms during the deposition of AlGaN section. This Al-rich shell[14,15], was observed in all our growth conditions (i.e. various metal/N ratio and growth temperature). Its formation is explained by the higher incorporation rate of Al at the sidewalls caused by the shorter diffusion length of the Al adatoms than that of Ga atoms.

Figure 1(b) shows the bright contrast of Fig. 1(a) taken at the NW center along the wire axis. Most of the investigated NWs show a *transition zone* [labeled (2)] between the GaN stem [labeled (1)] and the AlGaN section. The typical thickness of this area with a lower Al content is around ~10 nm. By reducing the growth temperature from 790°C to 735°C, the HAADF-STEM images show negligible changes in this zone in terms of chemical contrast and thickness. This fact indicates that the formation of this zone is not related to a surface reservoir of Ga atoms at the NW sidewalls, as Ga desorption/diffusion is highly temperature dependence in the range of our growth temperatures. After the *transition zone*, an Al-rich area [region (3)] was found, followed by an area displaying chemical ordering [region (4)]. The lengths of these two regions fluctuate from wire to wire. In the region (4), we usually find the 1:1 ratio ordering with the thickness of Al-rich and Ga-rich layers of about ~ 3 ML. However, it is difficult to determine precisely the Al content in both regions. The spontaneous chemical ordering is similar to what was observed in AlGaN[16,17,18] and InGaN films[19]. Theoretically, the ordering in wurtzite $Al_xGa_{1-x}N$ was suggested to be driven by the different binding energy of Al-N bond and Ga-N bond[20,21]. The adatom surface diffusion generally allows the incorporation of Al and Ga atoms at their different preferential sites, leading to the ordering sequence of Al-rich and Ga rich areas along the growth axis. Such a description is highly probable for the AlGaN growth in NWs where the contribution of adatom diffusion is significant. After region (4), several wires show the formation of a random alloy.

To understand the *transition zone* formation, we investigate an AlN/GaN NW superlattice by HAADF-STEM as shown in Fig. 2. Similarly, the image contrast reveals a *transition zone* with a length of ~ 8 nm at the first AlN/GaN interface [Label (1) in Fig. 2(a)]. The transition zones at the second, third and forth interfaces [Labels (2), (3) and (4)] are significantly shorter. From the image contrast, these regions should be Al-Ga intermixing areas. Besides, the intermixing process is seemingly more pronounced on the top of GaN rather than at the interface below them.

We propose that the intermixing at the AlN/GaN interface forms in order to partially relieve the strain induced by the lattice mismatch between AlN and GaN. We have explored the role of the strain via 3D strain simulations using NextNano[3]. The strain distribution was obtained by the minimization of the elastic strain energy through the application of zero-stress boundary conditions at the NW surface, which allows the NW to deform in all three spatial directions. In our case, the structure of interest is a 50 nm-diameter hexagonal GaN NW overgrown by various periods of 10-nm GaN/10 nm–AlN, i.e. the geometry corresponding to the structure in Fig. 2(a). We do not consider the formation of an Al shell as it is not observed in this HAADF-STEM image. The number of GaN/AlN periods in the simulation was increased sequentially, to assess the strain distribution at the topmost AlN/GaN interface after the deposition of each period. Note that the Al-Ga intermixing at the AlN/GaN interface was not taken into account for the simulation.

Figures 2(b) and (c) present the color map of the strain in $x$ direction ($e_{xx}$) in a hexagonal GaN NW capped with a single AlN layer. Figure 2(b) illustrates the in-plane distribution of tensile strain at the interface of GaN NW and AlN layer on the AlN side. The cross section in Fig. 2(c); shows the high degree of compressive strain (blue contrast) at the interface in the GaN region, and the tensile strain (red contrast) in the AlN region. From the simulation, the first AlN layer is strongly tensile strained by the GaN NW base. Figure 2(d) describes the elastic energy density at the topmost AlN/GaN interface on the AlN (■) side taken at the center of a GaN NW, plotted as a function of the number of AlN sections in the stack.

The stronger tensile strain in the first AlN layer probably enhances Al-Ga intermixing at the AlN/GaN interface. As the NW growth proceeds, the higher AlN barriers are less strained, thus giving lower driving force for the intermixing process, and consequently a shorter *transition zone*. The lower intermixing effect at the interface below each GaN layer than the one above them is possibly caused by the higher binding energy of the Al-N bond with respect to the Ga-N bond.

Figure 3 shows the PL spectra measured at 5 K and 300 K of GaN NWs with an $Al_xGa_{1-x}N$ section with nominal $x$= 0.21, 0.36 and 0.50 grown at 795°C. At 5 K, the PL peak which remains constant at around 3.47-3.5 eV is assigned to the GaN NW base, whereas those attributed to $Al_xGa_{1-x}N$ sections monotonically blue shift from 3.76 to 5 eV as a function of $x$. The full width at half maximum (FWHM) also increases from 78 meV to 445 meV when $x$ is increased to 0.50. These observed linewidths are larger than those reported in AlGaN 2D films[22]. The inset of Fig. 3(a) shows that the PL peak at 5K is lower and progressively deviates (up to 200 meV) from the expected bandgap value[23].

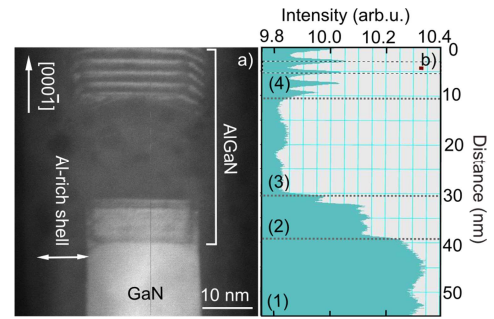

Figure 1 (a) HAADF-STEM image of a GaN NW with a single $Al_xGa_{1-x}N$ section. (b) Corresponding image contrast profile taken at the center of the NW shown in (a).

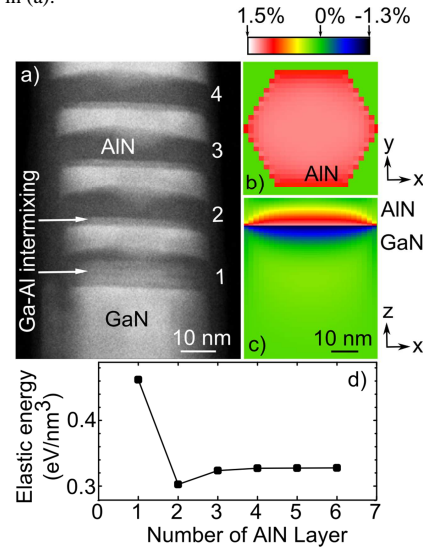

Figure 2 (a) HAADF-STEM image of a GaN NW with an AlN/GaN superlattice. The arrows indicate *the transition zone* at the AlN/GaN interfaces. (b,c) Contour plot $e_{xx}$ in a GaN NW with an AlN section on top: (b) in-plane view of the interface at the AlN side, and (c) cross-section at the center of GaN NW. (d) Plot of the elastic energy density measured at the topmost AlN/GaN interface on the AlN sides as a function of the number of AlN layers in the stack.

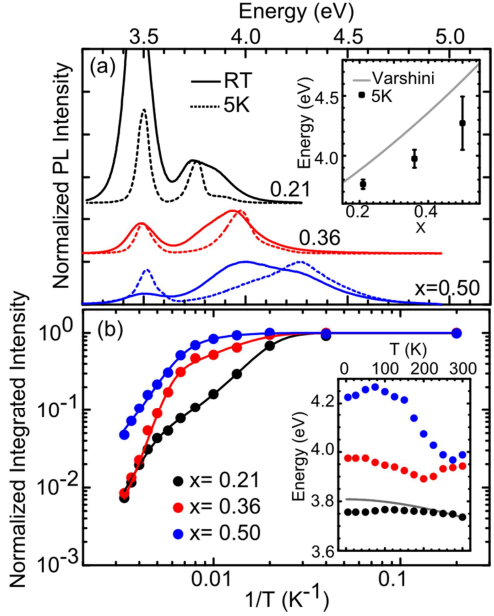

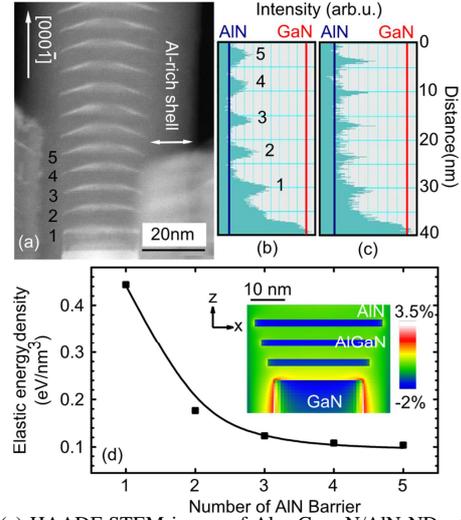

Figure 3 (a) PL spectra measured at 5 K and 300 K of GaN NWs with a 180 nm $Al_xGa_{1-x}N$ section with x= 0.21, 0.36 and 0.50. Inset: PL peak position as a function of $x$ in comparison with the expected bandgap value (solid line). The error bars represent the PL FWHM (b) Normalized integrated PL intensity of $Al_xGa_{1-x}N$ sections on GaN NWs as a function of temperature. Inset: Temperature dependent PL peak position. The solid line represents the shift expected from Varshni's equation

Figure 4 (a) HAADF-STEM image of $Al_{0.38}Ga_{0.62}N$/AlN NDs. (b) and (c) Image contrast taken at the center and at the edge of the NW in (a), respectively. (d) The plot of the elastic energy density taken at the center of the NW on the surface of the topmost 4-nm AlN barrier prior to the growth of the $Al_{0.38}Ga_{0.62}N$ ND, as a function of the number of AlN barriers. The inset shows the cross-section contour plot of $e_{xx}$ in the structure with 3 layers of 2-nm $Al_xGa_{1-x}N$ /4 nm AlN on GaN NW.

The integrated PL intensity of $Al_xGa_{1-x}N$ sections as a function of temperature where $x$ = 0.21, 0.36, and 0.50 is shown in Fig. 3(b). The intensity evolution as a function of temperature is described by an equation with two activation energies: $I(T) = I_0/(1+ae^{-E_{a1}/kT} +be^{-E_{a2}/kT})$, where $I$ is the PL intensity, $E_{a1}$ and $E_{a2}$ are the activation energies, $kT$ is the thermal energy, $a$ and $b$ are fitting parameters. The fits are represented by the solid lines in this figure. The extracted $E_{a1}$ in the range of 75-135 meV is attributed to the localization energy corresponding to the potential barriers which confine the carriers in Ga-rich area of $Al_xGa_{1-x}N$ section. The $E_{a2}$ with the value in the range of 15-20 meV could represent the activation energy accounting for the detrapping of excitons from the interface roughness fluctuation in the ordered alloy. Here, the internal quantum efficiency (IQE) is estimated by comparing the integrated intensity measured at room temperature with respect to that obtained at low temperature, assuming that IQE(5K) is 100%. The extracted IQE of AlGaN section increases from 0.7 to 5 % when $x$ increases.

The PL spectral shifts with temperature can provide information on the carrier localization in potential fluctuations in the alloy. The inset of Fig. 3(b) depicts the evolution of PL peak energy as a function of temperature, showing an *s*-shape behavior which deviates significantly from Varshni's equation[24]. This trend is explained by the exciton freeze-out in the local potential minima in the alloy at low temperature, followed by the onset of exciton thermalization with increasing temperature. The deviation becomes more pronounced with increasing $x$, similar to the case of AlGaN 2D films[25]. Potential fluctuations in ternary alloys can reduce the probability of carrier capture by nonradiative recombination centers. Therefore, the higher IQE in AlGaN section with larger Al content ($x<0.5$) could be associated to the larger alloy inhomogeneity.

To further enhance carrier localization in the AlGaN NW UV-emitters, $Al_xGa_{1-x}N$/AlN NDs were grown on the top of GaN NWs. From the growth conditions and the HAADF-STEM images, the thickness of the AlGaN NDs was found in the range of 1-4 nm and the thickness of the AlN barrier is around 4 nm. Figure 4 shows HAADF-STEM of a 25 nm diameter NW containing $Al_{0.38}Ga_{0.62}N$/AlN NDs. The ND thickness is the range of 2 nm, whereas that of the AlN barriers is ~4 nm. The diameter of the first $Al_xGa_{1-x}N$ ND is similar to the one of GaN NW stem. Due to the lateral growth of AlN barrier, the ND diameter progressively increases along the growth axis. Moreover, the shape of the NDs evolves to the pyramid-like shape with {1-103} facets, similar to GaN SK QDs[26]. From the image contrast, we found a radial compositional gradient, that is, the center of the ND has higher Al content than the edge area [Fig. 4(a)]. The contrast profile taken at the center of the NW in Fig. 4(b) reveals that the Al composition at the center of the NDs gradually increases along the growth axis (the image contrast become darker) and finally saturates at the third ND. In addition, the profile in Fig.4(c) taken at the edge of the NW shows that the Ga-rich area becomes better defined. This result suggests that the degree of lateral phase separation is more pronounced in the higher ND layer.

From our data, we suggest that that the strain can modify the shape, the axial and the radial distribution of Al composition of the AlGaN NDs. We performed 3D strain simulations of a 25 nm-diameter hexagonal GaN NW with various periods of 2 nm $Al_{0.38}Ga_{0.62}N$/ 4 nm AlN, to imitate our result shown in Fig 4(a). In this case, we consider the formation of the Al shell which encapsulates NDs and NW. The number of periods in the simulation was increased sequentially, to assess the strain distribution after the growth of an AlN barrier, following 0, 1, 2, 3 periods of $Al_{0.38}Ga_{0.62}N$/AlN ND. For simplicity, we assume random alloy in the AlGaN NDs. The inset of Fig. 4(d) presents the cross-section contour plot of $e_{xx}$ taken at the center of the NW after the growth of the 4$^{th}$ AlN barrier.

Figure 4(d) presents the summary of the elastic energy density taken at the surface of the topmost AlN barrier and at the center of the NW. The values are plotted as a function of the number of AlN barriers. The first 4-nm AlN barrier is highly tensile strained by the GaN NW base. Then, the strain gradually decreases and saturates after the fourth $Al_{0.38}Ga_{0.62}N$/AlN layer. We propose that the higher tensile strain in the first AlN barrier should facilitate Ga incorporation in the first $Al_{0.38}Ga_{0.62}N$ ND. As the NW growth proceeds, the tensile strain in the AlN barriers monotonically decreases, which enhances the Al incorporation at the center of the $Al_xGa_{1-x}N$ NDs. Radially, the efficient strain relaxation at the edge of the NDs possibly results in the higher accumulation of Ga atoms at that area.

Despite the chemical inhomogeneity in the $Al_xGa_{1-x}N$ NDs, we demonstrate that the emission wavelength can be tuned by varying the nominal Al content in the NDs and their thicknesses. Their CL spectra measured at 300K are presented in Fig. 5(a). By increasing $x$ from 0 to 0.38 and/or reducing the ND thickness from 4 to 1 nm, the CL peak emission can be adjusted from 350 nm to 240 nm, with a FWHM in the range of 50-500 meV. The broadening and structuration of the CL line in thinner NDs with higher Al content is attributed to the pronounced chemical inhomogeneity in the NDs due to the larger lattice mismatch with the GaN stem. The integrated CL intensity as a function of temperature of $Al_xGa_{1-x}N$ NDs with 1-2 nm thicknesses are shown in Fig. 5(b), in comparison to that of GaN QWs and 180 nm AlGaN/GaN NWs with similar Al content. The IQE of AlGaN NDs is in the range of 30-40%, significantly higher than that of 180 nm AlGaN sections and that of GaN QWs, which are around 5% and 0.5%, respectively.

In conclusion, we present the structural and optical studies of AlGaN sections and AlGaN/AlN NDs on GaN NWs. The Al-Ga intermixing at the Al(Ga)N/GaN hetero-interfaces in NWs and the chemical inhomogeneity in the $Al_xGa_{1-x}N$ NDs are associated to the strain relaxation process. The alloy inhomogeneity on nm length scales in AlGaN sections leads to potential fluctuations that induce carrier localization, which determines their optical behavior. Furthermore, it also causes the multi-peak spectral emission characteristic of AlGaN/AlN NDs. Despite the difficulty to control the alloy uniformity in AlGaN NDs, we demonstrate that the emission energy can be tuned from 240 to 350 nm with IQE around 30% by adjusting their thickness and Al content.

The authors would like to thank F. Donatini for technical assistance on CL measurement. We benefited from the access to the technological platform NanoCarac of CEA-Minatec for HAADF-STEM characterization. This work is partially supported by the French National Research Agency "UVLamp" project (ANR-2011 NANO-027), (#233950), JCJC project COSMOS (ANR-12-JS10-0002) and by the EU ERC-StG "TeraGaN" (#278428) project.

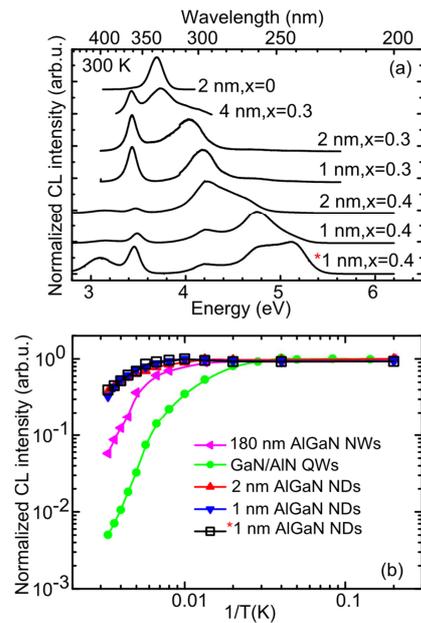

Figure 5 (a) CL spectra of GaN/$Al_xGa_{1-x}N$ NDs and $Al_xGa_{1-x}N$/AlN NDs grown with different $Al_xGa_{1-x}N$ ND thickness and Al contents. (b) Integrated CL intensity as a function of temperature of a 180 nm-$Al_{0.4}Ga_{0.6}N$ section, and 1 nm and 2 nm $Al_{0.4}Ga_{0.6}N$/AlN NDs, in comparison to that of GaN/AlN QWs. Note that * represents $Al_xGa_{1-x}N$/AlN NDs which are grown in the N-rich regime instead of the N-limited regime.